\documentclass[12pt,twoside]{article}

\pdfoutput=1

\usepackage[usenames]{color}
\usepackage{lscape}
\usepackage{amssymb}
\usepackage{rotating}
\usepackage[T1]{fontenc}
\usepackage{graphicx}
\usepackage{fancyhdr}
\usepackage{amsmath}
\usepackage{ae}
\usepackage{aecompl}
\usepackage{hyperref}

\def\pa{\partial}

\def\ov{\overline}

\def\BR{\rm{ BR}}

\newcommand{\be}{\begin{equation}}
\newcommand{\ee}{\end{equation}}
\newcommand{\ba}{\begin{eqnarray}}
\newcommand{\ea}{\end{eqnarray}}

\newlength{\dinwidth}
\newlength{\dinmargin}
\begin{document}

\thispagestyle{empty}

\begin{flushright}
 EFI-16-04
\end{flushright}

\vspace*{1cm}

\centerline{\Large\bf Enhancing the Higgs associated production} 
\centerline{\Large\bf with a top quark pair}

\vspace*{15mm}

\centerline{Marcin Badziak${}^a$ and Carlos E.~M.~Wagner$^{b,c,d}$}
\vspace*{5mm}

\centerline{${}^a$\em Institute of Theoretical Physics,
Faculty of Physics, University of Warsaw} 
\centerline{\em ul.~Pasteura 5, PL--02--093 Warsaw, Poland} 

\centerline{$^b$\em Enrico Fermi Institute, University of Chicago, Chicago, IL 60637, USA}

\centerline{$^c$\em High Energy Physics Division, Argonne National Laboratory, Argonne, IL 60439, USA}

\centerline{$^d$\em Kavli Institute for Cosmological Physics, University of Chicago, Chicago, IL 60637, USA}

\vskip 1cm

\centerline{\bf Abstract}

It is pointed out that in a wide class of models reminiscent of type-II Two-Higgs-Doublet Models (2HDM) the signal of the Higgs produced in association with a
top-antitop quark pair ($tth$) and decaying into gauge bosons can be significantly larger than the Standard Model (SM) prediction without violating any
experimental
constraints. The crucial feature of these models is enhanced (suppressed) Higgs coupling to top (bottom) quarks and existence of light colored particles that
give negative contribution to the effective Higgs coupling to gluons resulting in the gluon fusion rates in the gauge boson decay channels close to SM
predictions. We demonstrate this mechanism in NMSSM with light stops and show that $tth$ signal in the $WW$ decay channel can be two times larger than
the SM prediction, as suggested by the excesses observed by ATLAS and CMS, provided that the Higgs-singlet superpotential coupling $\lambda\gtrsim0.8$ and
the MSSM-like Higgs boson masses are in the range of 160 to 300 GeV.  

\vskip 3mm

\newpage

\section{Introduction}

The most important legacy of the first run of the LHC is the discovery of a 125 GeV Higgs boson~\cite{Higgsdiscovery}. The measured Higgs signal
rates in all channels agree with the SM prediction at $2\sigma$ level~\cite{Higgsdata}. Moreover, in the most precisely measured channels, such as 
gluon fusion ones with a Higgs decaying into gauge bosons, the agreement is typically at the $1\sigma$ level.
One of the main goals of the 13 TeV LHC is to improve measurements of the Higgs properties. However, the possibility of extracting information on new physics from measurements of 
Higgs rates in the gluon fusion production channels is somewhat limited by systematics and the theoretical uncertainty of the SM gluon fusion production
cross-section~\cite{gluonfusionSM}.
One can then naturally ask whether there are better channels for the discovery of  New Physics from  Higgs measurements at the LHC. Fortunately, the rate measurements in some channels are currently statistically limited and can benefit a lot from the high luminosity expected to be delivered by the 13 TeV LHC. Among
these channels, a  particularly interesting one is the Higgs production  in association with a
top-antitop quark pair ($tth$). 

The top quark is often considered as a window to New Physics. This statement is supported by the fact that the top quark mass is much larger than all other
quarks and its SM Yukawa coupling is of order unity. In consequence, there are many phenomena involving quarks that are measured (or can be
measured in a near future) only for top quarks. That said, it should
be emphasized that the top quark Yukawa coupling has not been measured directly so far. The only hint that the top  quark Yukawa coupling is indeed very close to
the SM prediction comes from the measurements of the Higgs gluon fusion production rates that agree very well with the SM. In the SM, the gluon
fusion production cross-section is to a large extent controlled by the top quark Yukawa coupling. However, in many extensions of the SM there are
new coloured particles that can contribute to the gluon fusion cross-section, interfering with the top quark loop. In such a case, simple relation between
the top quark Yukawa coupling and the gluon fusion production cross-section is lost.\footnote{The degeneracy in the gluon fusion production cross-section between the top quark and New Physics contributions can be broken by studying production of a boosted Higgs with a jet \cite{Azatov:2013xha,Grojean:2013nya}.}
 Therefore, in general   
it is the $tth$ production which may give access to the top quark Yukawa coupling directly. 

A particularly interesting and timely question at the dawn of the 13 TeV LHC run is whether the $tth$ signal rates can be substantially enhanced with
respect to the SM prediction. If a big enhancement is indeed realized in Nature we should discover it at the LHC run two. Moreover, the LHC data
from the first run give some hints for such enhancement since a fit to the combined ATLAS and CMS data yields  a signal strength
\begin{equation}
\mu^{\rm tth}=2.3^{+0.7}_{-0.6},
\end{equation}
for
the $tth$ production cross-section normalised to the SM prediction \cite{Higgsdata}. Many different final states contribute to this enhancement, both
at ATLAS \cite{tthATLAS} and CMS \cite{tthCMS}, but the most significant excesses are observed in multilepton final states which probe mainly the
$tth$ production in the $WW$ decay channel. For the $\gamma\gamma$ channel the central values are also above the SM prediction in both experiments,
with particularly large enhancement observed at CMS. 
All of the above suggests enhancement of $tth$ signal rates with a Higgs decaying into gauge bosons. 

During the last year, there have been several analyses that interpreted the excess in the $tth$ searches in New Physics models. Most of those works
focused on the same-sign dilepton excess in the $tth$ searches and interpreted it as a signature of a new particle, see
e.g.~Refs.~\cite{Huang:2015fba,Chen:2015jmn}. To the best of our knowlegde, only Ref.~\cite{Angelescu:2015kga} interpreted the $tth$ excess in a model
with enhanced Higgs coupling to top quarks.

In the present paper we show that the $tth$ production rate with a Higgs decaying into gauge boson can be more than a factor of two larger than in
the SM without violating any existing data in a wide class of models reminiscent of type-II Two-Higgs-Doublet Models (2HDM). In order to achieve this, the
existence of new light coloured particles is necessary to disentangle the top quark Yukawa coupling from the effective Higgs coupling to gluons. We
demonstrate this effect using stops as an example which, if sufficiently light and highly mixed, can reduce the effective Higgs coupling to gluons
keeping gluon fusion rates close to the SM prediction when the $tth$ production channel is enhanced. We also show that such a big $tth$ enhancement
can be accommodated in the Next-to-Minimal Supersymmetric Standard Model (NMSSM) \cite{reviewEllwanger} if the Higgs-singlet superpotential coupling
$\lambda$ is large enough
and the MSSM-like Higgs bosons are in the range of several hundreds of GeV.

The rest of the paper is organized as follows. In section \ref{sec:2HDM} we study the $tth$ signal rates in type-II 2HDM and show that its possible
enhancement is very limited by the Higgs data in the gluon fusion production channels. In section \ref{sec:2HDMstops} we add light stops to type-II
2HDM and show that large $tth$ enhancement can be consistent with the experimental data. In section \ref{sec:NMSSM} we show that such enhancement is
possible in NMSSM and discuss implications for the spectrum of MSSM-like Higgses taking into account experimental constraints and present several
benchmark points. We summarize our results in section \ref{sec:concl}. 

\section{tth in type-II 2HDM}
\label{sec:2HDM}

Let us start with an analysis of type-II 2HDM which mimics certain regions of the MSSM, as well as the NMSSM with decoupled singlet. The tth production cross-section is
controlled by the top quark Yukawa coupling. Since the SM Higgs production cross-sections are computed with better precision than in any of the SM extensions we focus
on the tth production cross-section normalised to the SM prediction:
\begin{equation}
\sigma^{tth}\equiv \frac{\sigma(gg\to tth)}{\sigma^{SM}(gg\to tth)}=c_t^2  \,,
\end{equation}
where $c_t$ is the top quark Yukawa coupling normalised to its SM value. 

The LHC experiments measure the production cross-section times branching ratio so it is
useful to define theoretically predicted signal strengths modifiers as:
\begin{equation}
 R_i^j\equiv\frac{\sigma^j\times {\rm BR}(h \to i)}{\sigma^{j{\rm SM}}\times {\rm BR}^{\rm SM}(h \to i)} \,.
\end{equation}
Throughout the paper, we distinguish the theoretical predictions for the signal strengths from the corresponding LHC measurements, that we define in the conventional way as $\mu_i^j$. In the present case $R_i^j$ depend on the Higgs couplings to up-type
fermions $c_t$, down-type fermions $c_b$ and massive gauge bosons $c_V$, as well as on effective Higgs couplings to gluons and photons that depend on
the SM
couplings and may receive contributions from New Physics. Formulae for $R_i^j$ as a function of these couplings are given in the Appendix.   

\begin{figure}
 \includegraphics[width=0.44\textwidth]{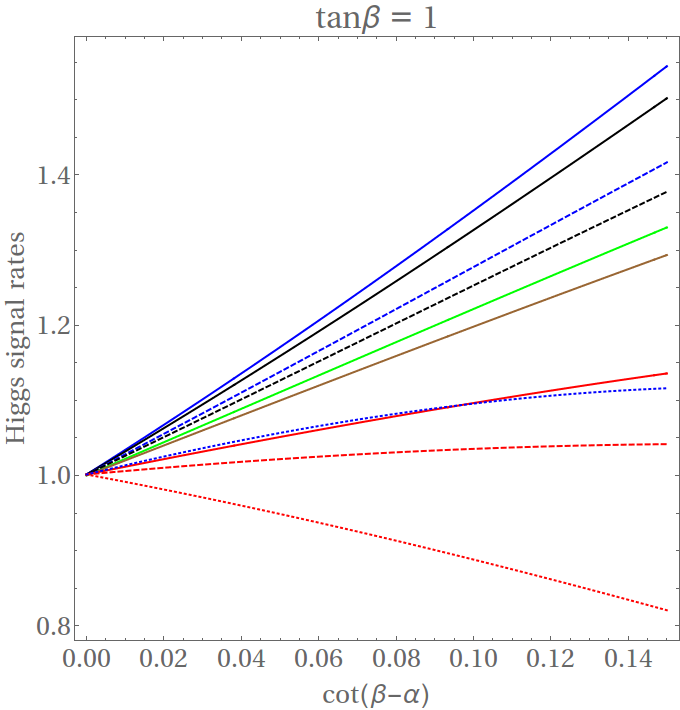}
  \includegraphics[width=0.44\textwidth]{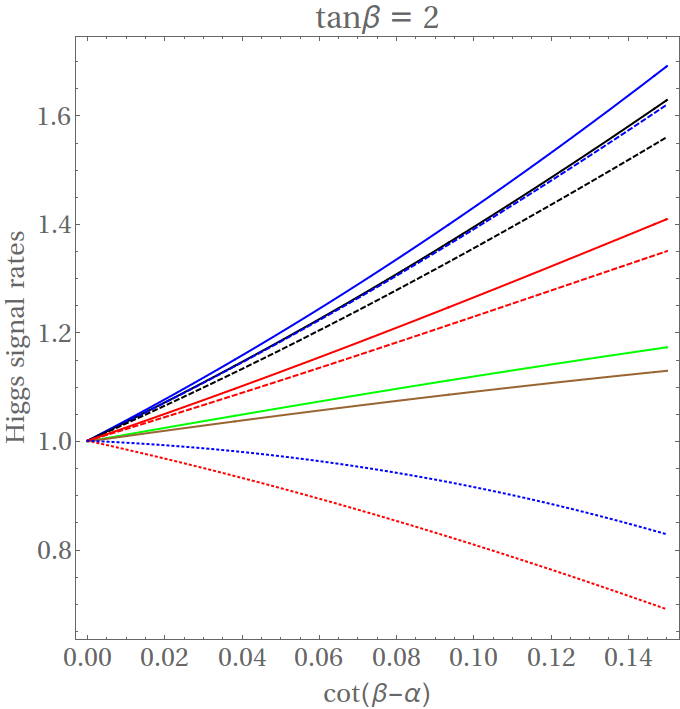}
    \includegraphics[width=0.1\textwidth]{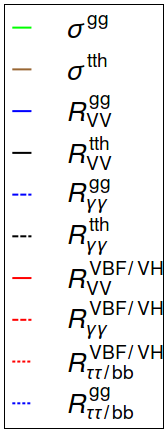}
\caption{Dependence of Higgs signal rates on $\cot\left(\beta-\alpha\right)$ for $\tan\beta=1$ (left) and 2 (right) in type-II 2HDM. }
\label{fig:tth_2HDM}
\end{figure}

In the type-II 2HDM the couplings (normalised to SM) read:
\begin{align}
\label{eq:ct}
&c_t=\frac{\cos\alpha}{\sin\beta}=\sin\left(\beta-\alpha\right) + \cot\beta \cos\left(\beta-\alpha\right) \,, \\
\label{eq:cb}
 &c_b=-\frac{\sin\alpha}{\cos\beta}=\sin\left(\beta-\alpha\right) - \tan\beta \cos\left(\beta-\alpha\right) \,, \\
 \label{eq:cV}
 &c_V=\sin\left(\beta-\alpha\right) \,,
\end{align}  
The SM couplings are obtained in the decoupling limit $\alpha=\beta-\pi/2$. It is clear from the above formulae that significant deviations from the SM for the
$tth$ production cross-section can only occur for small values of $\tan\beta$ and away from the decoupling limit. This generically implies relatively small mass of additional Higgs bosons,  especially in weakly-coupled models of new physics where $\cos\left(\beta-\alpha\right)\sim M_Z^2/m_H^2$ is typically expected. It is important to note the anti-correllation
between
$c_t$ and $c_b$. If one is enhanced, the other one is suppressed and vice-versa. Moreover, for $\tan\beta>1$ the bottom Yukawa coupling deviates from the SM
more than the top  quark Yukawa. This is particularly important since the bottom Yukawa coupling controls to large extent the total decay width of the Higgs because
the SM Higgs branching ratio to bottom and tau pairs exceeds in total 60\%. Therefore, all the branching ratios strongly deviate from the SM prediction if
$c_b$ strongly deviates from $c_V$. Since the LHC Higgs measurements are close to the SM predictions this puts strong constraint on possible deviations of $c_t$
from one.  

The dependence of $\sigma^{tth}$ and other rates on $\cot\left(\beta-\alpha\right)$ for $\tan\beta=1$ and $2$ is  shown in Fig.~\ref{fig:tth_2HDM}.
Due to the observed excess in $\mu^{\rm tth}_{WW}$, it is particularly interesting to investigate predictions for $R^{\rm tth}_{VV}$, where $V=W$ or
$Z$. It can be seen
from 
eqs.~\eqref{eq:ct}-\eqref{eq:cb}  that in type-II 2HDM $R^{\rm tth}_{VV}$ can be enhanced only for
$\cot\left(\beta-\alpha\right)>0$. As is shown in Fig.~\ref{fig:tth_2HDM}, in such a case, both the $tth$ production cross-section and the branching ratio to $WW$ is enhanced. However, a large
enhancement
of $R^{\rm tth}_{VV}$ is constrained by the existing LHC Higgs data which in most cases agree quite well with the SM predictions. For easy comparison we reproduce the result of the fit to the combined ATLAS and CMS data in Table \ref{tab:Higgsdata}. The main constraint
comes from
the measurements of $R^{\rm gg}_{VV}$ which is even slightly bigger than  $R^{\rm tth}_{VV}$ because the enhancement of the gluon-fusion cross section becomes bigger than the
one of the $tth$
cross-section when the $hb\ov{b}$ coupling is suppressed, cf. eqs.~\eqref{eq:hatcg} and \eqref{eq:cgBSM}. 

\begin{table}[t]
\centering
\begin{tabular}{c|c}
Channel & ATLAS+CMS combined result    \\
\hline
\hline
$\mu^{\rm gg}_{\gamma\gamma}$     & $1.19^{+0.28}_{-0.25}$     \\
$\mu^{\rm gg}_{ZZ}$               & $1.44^{+0.38}_{-0.34}$    \\
$\mu^{\rm gg}_{WW}$               & $1.00^{+0.23}_{-0.20}$    \\
$\mu^{\rm gg}_{\tau\tau}$               & $1.10^{+0.61}_{-0.58}$    \\
$\mu^{\rm gg}_{bb}$     & $1.09^{+0.93}_{-0.89}$     \\
$\mu^{\rm VBF/VH}_{\gamma\gamma}$     & $1.05^{+0.44}_{-0.41} $    \\
$\mu^{\rm VBF/VH}_{ZZ}$               & $0.48^{+1.37}_{-0.91}$    \\
$\mu^{\rm VBF/VH}_{WW}$               & $1.38^{+0.41}_{-0.37} $   \\
$\mu^{\rm VBF/VH}_{\tau\tau}$               & $1.12^{+0.37}_{-0.35} $   \\
$\mu^{\rm VBF/VH}_{bb}$     & $0.65^{+0.30}_{-0.29} $   \\ 
\end{tabular}
\caption{Observed Higgs signal strengths from the combination of the ATLAS and CMS data, corresponding to Table 13 of ref.~\cite{Higgsdata}. }.
\label{tab:Higgsdata}
\end{table}

We conclude that in type-II 2HDM, without the addition of new particles, it is not possible to strongly enhance $R^{\rm tth}_{VV}$ while keeping other rates in a good agreement with the SM
predictions. 

\section{tth in type-II 2HDM with light stops}
\label{sec:2HDMstops}

The conclusion of the previous section would not hold if there existed new coloured states that modify gluon-fusion production cross-section. Such modification of
effective coupling of the Higgs to gluons  is parameterised by $\delta c_g$ in our computation of the cross sections and branching ratios given in  eq.~\eqref{eq:cgBSM}. In this paper we focus on light stops as a source of
$\delta c_g$ because the Higgs sector of minimal SUSY models reduces to the class of Type-II 2HDM in certain limits. Nevertherless, one should keep in mind that modification of $c_g$ can
originate from other light coloured states, see e.g.~Ref.\cite{Falkowski}, so the mechanism we present is applicable more generally. 
  
Type-II 2HDM with light stops that we consider should be thought of a simplified model of an extended model which reduces to the MSSM at low energies. One example that we shall analyze below is the  NMSSM in which the singlet is decoupled and does not effectively mix
with the Higgs doublets. Note that an ultraviolet completion to the MSSM is needed because for small $\tan\beta$ light stops cannot account for the 125 GeV Higgs mass.

Light stops modify the effective Higgs coupling to gluons and photons in the following way, see e.g.~Refs.\cite{Falkowski,Djouadi}:
\begin{equation}
 \frac{c_g}{c_g^{\rm SM}}=\frac{c_{\gamma}}{c_{\gamma}^{\rm SM}}=c_t + \frac{m_t^2}{4}\left[c_t\left( \frac{1}{m_{\tilde{t}_1}^2} +
\frac{1}{m_{\tilde{t}_2}^2}\right) - \frac{\tilde{X}_t^2}{m_{\tilde{t}_1}^2 m_{\tilde{t}_2}^2} \right] \,,
\label{eq:cg_stopeff}
\end{equation}
where $\tilde{X}_t^2 \equiv X_t  \left( A_t \frac{\cos\alpha}{\sin\beta} + \mu \frac{\sin\alpha}{\sin\beta} \right)$ with the stop mixing parameter given by $X_t \equiv A_t-\mu/\tan\beta$ (note: in the decoupling limit $\tilde{X}_t^2 = X_t^2$). 
In the above formula the corrections of order $\mathcal{O}(m_h/m_{\tilde{t}})$ are neglected because they have very small impact on the results already for stop
masses of about 200 GeV. We also neglect the NLO QCD corrections which have a rather small effect on the results \cite{Djouadi,NSUSYfits}. 

\begin{figure}
 \includegraphics[width=0.44\textwidth]{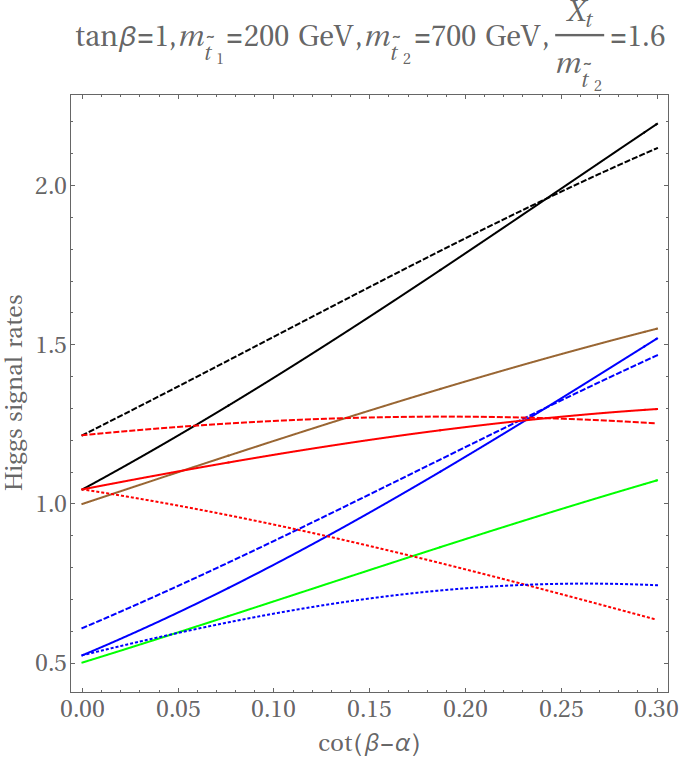}
  \includegraphics[width=0.44\textwidth]{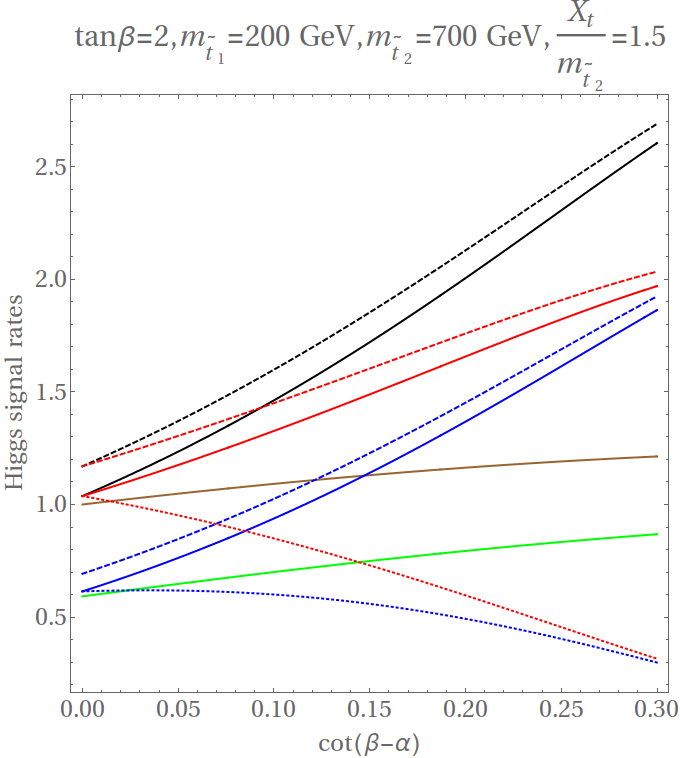}
  \includegraphics[width=0.1\textwidth]{tth_2HDM_legend5.png}
\caption{Dependence of the Higgs signal rates on $\cot\left(\beta-\alpha\right)$ for $\tan\beta=1$ and $2$  in type-II 2HDM with light
stops. }
\label{fig:tth_2HDMstops}
\end{figure}

In order to enhance the $tth$ production channel keeping the gluon fusion rates close to its SM values the effective Higgs coupling to gluons must be
smaller
than the Higgs coupling to top quark. It should be clear from eq.~\eqref{eq:cg_stopeff} that
for relatively large stop mixing parameters,  $\frac{\tilde{X}_t^2}{m_{\tilde{t}_2}^2}\gtrsim c_t$,  the modification of the gluon coupling $c_g/c_g^{\rm SM}$ can be smaller than $c_t$. In this cases $R^{\rm gg}_{VV} < R^{\rm tth}_{VV}$, as required by data.  In
the left panel of Fig.~\ref{fig:tth_2HDMstops} we show an example with stop masses of 200 and 700 GeV and $\tan\beta=1$. As can be seen from this
figure, values of $R^{\rm
tth}_{VV}$
of about 2 are
possible while keeping $R^{\rm gg}_{VV}$ and $R^{\rm gg}_{\gamma\gamma}$ only 30\% above the SM prediction, which is within the present 1$\sigma$ experimental
bounds for these Higgs production channels \cite{Higgsdata}, see also point B1 in Table~\ref{tab:benchmarks2HDM}. Notice also that for $R^{\rm
tth}_{VV}\approx2$ the Higgs $tth$ production cross-section $\sigma^{\rm tth}$ is enhanced by about 45\% while the rest of the enhancement originates
from suppressed $hb\bar{b}$ coupling that results in enhanced BR$(h\to VV)$. Another consequence of suppressed
$hb\bar{b}$
coupling are suppressed Higgs decays to $b\bar{b}$ and $\tau\tau$. Nevertheless, for $\tan\beta=1$ the signal strengths in these decay channels are about 0.75
(in gluon fusion production mode, as well as in the Higgs associated production with a weak boson (VH) and weak boson fusion (VBF) production
channels). Such small suppression is even preferred by the current LHC measurements of the $b\bar{b}$ decay channel.
Similar suppression is not observed in the $\tau\tau$ decay channel but values of $R^{\rm VBF/VH}_{\tau\tau}$  as low as about 0.4 are allowed at
$2\sigma$ level for the VBF/VH  production channel. The gluon fusion rate in the $\tau\tau$ channel is poorly measured and even zero is allowed at $2\sigma$ level.

As $\tan\beta$ increases, suppression of the $hb\bar{b}$ coupling becomes stronger while the enhancement of the $ht\bar{t}$ coupling becomes weaker. In
consequence, enhancement of $R^{\rm tth}_{VV}$ is mainly driven by enhancement of BR$(h\to VV)$. This is demonstrated for $\tan\beta=2$ in the right
panel of
Fig.~\ref{fig:tth_2HDMstops}. In this case $R^{\rm tth}_{VV}=2$ is obtained with $\sigma^{\rm tth}$ only 20\% above the SM prediction. This results in
larger
deviations of other signal rates from the SM predictions. The gluon fusion production rate in the gauge bosons decay channel is not an issue because
it can be adjusted to SM-like values by appropriate choice of $X_t/m_{\tilde{t}_2}$. The gluon fusion rate in the $\tau\tau$ turns out to be quite
low but it poses no tension with the
current LHC data. Constraints from the VBF/VH production channels are more important since these channels are not affected by presence of light
stops. VH is the most relevant production channel for $h\to b\bar{b}$ while for $h\to \tau\tau$ this is VBF. As long as $\tan\beta\lesssim1.5$,
$R^{\rm VBF/VH}_{\tau\tau}$ sets the strongest upper limit on $R^{\rm tth}_{VV}$.

\begin{figure}[t]
\centering
 \includegraphics[width=0.5\textwidth]{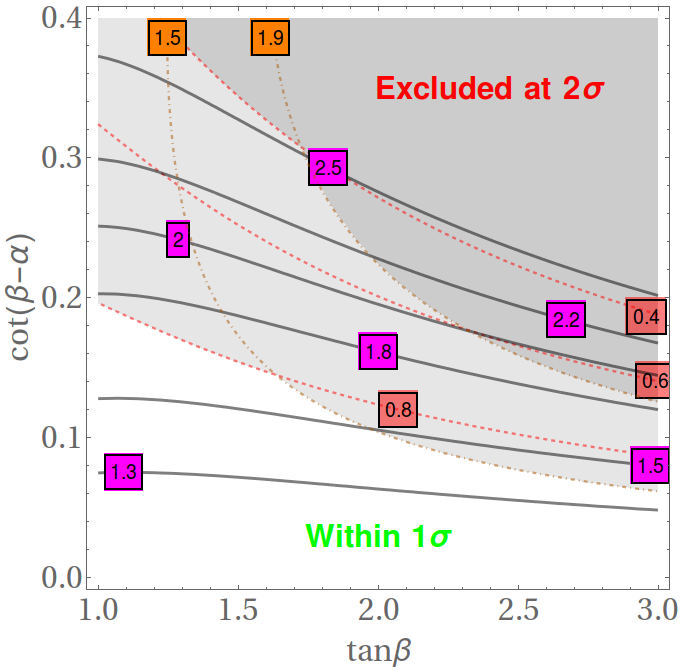}
\caption{Contour plot of $R^{\rm tth}_{VV}$ (black solid lines with magenta labels), $R^{\rm VBF/VH}_{\tau\tau}$ (dashed red lines with red labels)
and $R^{\rm VBF/VH}_{\gamma\gamma}$ (dot-dashed orange lines with orange labels) in the plane ($\tan\beta$, $\cot(\beta-\alpha)$) in type-II 2HDM
with light stops. Darker grey region is excluded at $2\sigma$ by at least one channel, while in the white region all the rates
are within 1$\sigma$ from the corresponding central values. The value of the gluon fusion rates can be always adjusted by a proper choice of
parameters
in the
stop sector. In order to calculate the total decay width $\delta c_g=-0.25$ is used in this plot, which is a typical value needed to keep the gluon
fusion rates close to the SM values when $tth$ rates are enhanced, while $\delta c_{\gamma}=2\delta c_g/9$. The position of the contours vary rather
mildly with
$\delta c_g$.}
\label{fig:tth_2HDMcontour}
\end{figure}

For the Higgs
decaying to gauge bosons VH and VBF
channels are measured much less precisely than the gluon fusion one. Nevertheless, for $\tan\beta\gtrsim1.5$ these channels start to compete with
$R^{\rm VBF/VH}_{\tau\tau}$ in setting an upper limit on possible enhancement of $R^{\rm tth}_{VV}$, as can be seen from
Fig.~\ref{fig:tth_2HDMcontour} and Table~\ref{tab:benchmarks2HDM} with benchmark points.
Currently the strongest upper limit on signal rates in these production channels is about 1.9 (1.5) at $2\sigma$ (1$\sigma$) for $R^{\rm
VBF/VH}_{\gamma\gamma}$. Moreover,  if the gluon fusion rate is suppressed by light stops then $\Gamma(h\to \gamma\gamma)$ is enhanced which makes
this channel even more important. Nevertheless, for $\tan\beta=2$  it is still possible to obtain
$R^{\rm tth}_{VV}\sim2$ while keeping other rates within $2\sigma$ from the experimental
central values. For large enough $\tan\beta$,  when the  enhancement of the $ht\bar{t}$
coupling becomes small, $R^{\rm
VBF/VH}_{\gamma\gamma}$ becomes bigger than $R^{\rm
tth}_{VV}$. This happens for $\tan\beta\gtrsim2.5$, as can be seen from Fig.~\ref{fig:tth_2HDMcontour}.

A preference for low $\tan\beta$ is emphasized in Fig.~\ref{fig:tth_2HDMcontour}. It can be seen that for $1\lesssim\tan\beta\lesssim1.5$, $R^{\rm
tth}_{VV}$ can exceed 2.5, while keeping other rates within $2\sigma$  from the corresponding experimental central values. 
In order to keep all the rates  within $1\sigma$, $R^{\rm VBF/VH}_{\tau\tau}$ must be above about 0.8 which for $\tan\beta=1$ allows
for $R^{\rm tth}_{VV}$ up to about 1.8.

 It is interesting to note that maximal value of $R^{\rm tth}_{VV}$, consistent with other
data at $2\sigma$, decreases quite slowly with $\tan\beta$. The reason is that  the branching
ratio of the Higgs decaying to $VV$ increases with $\tan\beta$ which partly compensates the decrease of $\sigma^{\rm tth}$.  
Keeping all the rates within $2\sigma$  from the corresponding experimental central values, $R^{\rm
tth}_{VV}=2$ is possible as long as $\tan\beta\lesssim2.5$, even if $R^{\rm VBF/VH}_{\tau\tau} \geq 0.6$ is taken, which seems to be more realistic
than allowing values as low as 0.4 for this quantity.

Let us end this section with a comment that in generic supersymmetric extensions of the SM there is a correlation between the Higgs couplings and the Higgs mass so typically one expect additional constraints on possible $tth$ enhancement imposed by the Higgs mass measurement of 125 GeV. Moreover, light highly-mixed stops required to keep the gluon fusion rate under control may induce non-negligible loop corrections to the off-diagonal entry of the Higgs mass matrix, hence also to the Higgs couplings, especially if the second Higgs doublet is light.  In particular, this is the case for NMSSM which we discuss in detail in the next section.

\begin{table}[t]
\centering
\begin{tabular}{c|ccc}
& {\rm B1} & {\rm B2} & {\rm B3}    \\
\hline
$\tan \beta$ & 1 & 1.5 & 2  \\
$\cot \left(\beta-\alpha\right)$ & 0.25 & 0.22 & 0.18  \\
\hline
\hline
$m_{\tilde{t}_1}$ & 200 & 200 & 210   \\
$m_{\tilde{t}_2}$ & 700 & 700 & 700   \\
$\tilde{X}_t/m_{\tilde{t}_2}$ & 1.7 & 1.6 & 1.6 \\
\hline
$R^{\rm tth}_{VV}$              & 2.02 & 1.96 & 1.90    \\
$R^{\rm tth}_{\gamma\gamma}$    & 2.09 & 2.09 & 2.07    \\
$R^{\rm gg}_{VV}$               & 1.18 & 1.21 & 1.19    \\
$R^{\rm gg}_{\gamma\gamma}$     & 1.22 & 1.29 & 1.29    \\
$R^{\rm VBF/VH}_{VV}$           & 1.29 & 1.49 & 1.60    \\
$R^{\rm VBF/VH}_{\gamma\gamma}$ & 1.33 & 1.59 & 1.74   \\
$R^{\rm VBF/VH}_{\tau\tau}$     & 0.73 & 0.67 & 0.66    \\
\end{tabular}
\caption{List of benchmark points for Type-II 2HDM with light stops. All masses are in GeV. }.
\label{tab:benchmarks2HDM}
\end{table}

\section{tth in the NMSSM}
\label{sec:NMSSM}

Let us now discuss $tth$ production in NMSSM which is a more restrictive framework because the mixing angles in the Higgs sector are functions of
NMSSM parameters
which cannot take arbitrary values. We focus on the general NMSSM for which the MSSM superpotential is supplemented by (we use the notation of
ref.~\cite{nmssmmixing}):
\begin{equation}
\label{W_NMSSM}
 W_{\rm NMSSM}= \lambda SH_uH_d + f(S) \,.
\end{equation}   
The first term is the source of the effective higgsino mass parameter, $\mu_{\rm eff}\equiv\lambda v_s$ (we drop the subscript ``eff'' in the rest of the
paper), while the second term parametrizes various versions of NMSSM. In the simplest version, known as the scale-invariant NMSSM, $f(S)\equiv\kappa S^3/3$,
while in more general models $f(S)\equiv\xi_F S+\mu'S^2/2+\kappa S^3/3$.

It is more convenient for us to work in the  Higgs basis $(\hat{h}, \hat{H}, \hat{s})$, where $\hat{h}=H_d\cos\beta + H_u\sin\beta$, $\hat{H}=H_d\sin\beta -
H_u\cos\beta$ and $\hat{s}=S$. This is because $\hat{h}$ field has exactly the same couplings to the gauge bosons and fermions as the SM Higgs field. The field
$\hat{H}$ does not couple to the gauge bosons and its couplings to the up and down fermions are the SM Higgs ones rescaled by $\tan\beta$ and $-\cot\beta$,
respectively. The mass eigenstates are denoted as $s$, $h$, $H$, with the understanding that $h$ is the SM-like Higgs.  

In the hatted basis the tree-level Higgs mass matrix in general NMSSM is given by:
\begin{equation}
 \hat{M}^2=
\left(
\begin{array}{ccc}
  \hat{M}^2_{hh} & \hat{M}^2_{hH} & \hat{M}^2_{hs} \\[4pt]
   \hat{M}^2_{hH} & \hat{M}^2_{HH} & \hat{M}^2_{Hs} \\[4pt]
   \hat{M}^2_{hs} & \hat{M}^2_{Hs} & \hat{M}^2_{ss} \\
\end{array}
\right) \,,
\end{equation}
where, at tree level, 
\begin{align}
\label{Mhh}
 &\hat{M}^2_{hh} = M_Z^2\cos^2\left(2\beta\right)+ \lambda^2 v^2\sin^2\left(2\beta\right) \,, \\
\label{MHH}
&\hat{M}^2_{HH} = (M_Z^2-\lambda^2 v^2)\sin^2\left(2\beta\right) + \frac{2 B \mu}{\sin\left(2\beta\right)} \,, \\
\label{Mss}
&\hat{M}^2_{ss}=  \frac{1}{2} \lambda v^2 \sin2\beta \left(\frac{\Lambda}{v_s}- \langle\pa^3_S f\rangle\right) + \Upsilon \,,\\
\label{MhHfull}
& \hat{M}^2_{hH} = \frac{1}{2}(M^2_Z-\lambda^2 v^2)\sin4\beta \,, \\
\label{Mhs}
& \hat{M}^2_{hs} =  \lambda v (2\mu-\Lambda \sin2\beta) \,, \\
\label{MHs}
& \hat{M}^2_{Hs} = \lambda v \Lambda \cos2\beta \,.
\end{align}
where  $\Lambda\equiv A_{\lambda}+\langle\pa^2_S f\rangle$,
$B\equiv A_{\lambda}+ \langle\pa_S f\rangle/v_s+m_3^2/(\lambda v_s)$, $\Upsilon\equiv\langle(\pa^2_S f)^2\rangle + \langle\pa_S f \pa^3_S f\rangle -
\frac{\langle\pa_S f \pa^2_S f\rangle}{v_s} + A_{\kappa} \kappa v_s - \frac{\xi_S}{v_s}$ and $v\approx174$ GeV.

Since we are mainly interested in the enhancement of the $tth$ production cross-section small mixing between the Higgs $h$ and the singlet is preferred.  Since the main effects come from admixture of the $h$ and $H$, we
assume that the singlet components of $h$ and $H$ are negligible, which can be obtained by taking appropriately large $\hat{M}^2_{ss}$.
Nevertheless, even with approximately decoupled singlet NMSSM is very different from MSSM because of the Higgs-singlet interaction
controlled by the coupling $\lambda$.   For instance, as was discussed in Ref.~\cite{alignment} the mixing between $h$ and $H$ take small values for $\lambda \simeq 0.6$--0.7, leading to an effective alignment of the SM-like Higgs bosons for these values of the trilinear couplings. 

These properties may be easily understood by studying the CP-even Higgs mass matrix properties.  For values of $\tan\beta$ of order one, the dominant
loop correction contributes to $M^2_{H_uH_u}$ entry but after the rotation to the Higgs basis gives also correction to the diagonal and
off-diagonal entries of the CP-even Higgs mass matrix (for the approximate expression of these corrections, see, for instance, Ref.~\cite{alignment}).
 We shall parametrize these corrections by those affecting the $hh$ matrix element,
\begin{equation}
\hat{M}_{hh}^2 = M_Z^2 \cos^2(2\beta) + \lambda^2 v^2 \sin^2(2\beta) + \Delta_{\rm loop}^2
\end{equation}
 It is straightforward to show that in this case
\begin{equation}
\label{eq:cot_betaalpha_NMSSM}
\cot(\beta-\alpha)=\frac{\frac12(M^2_Z-\lambda^2 v^2)\sin4\beta-\Delta_{\rm loop}^2/\tan\beta}{\hat{M}^2_{HH}-m_h^2}
\end{equation} 
where we used the notation of the 2HDM which is justified as long as the singlet admixture in $h$ and $H$ is negligible.  
The enhancement of $R^{\rm tth}_{VV}$ requires $\cot\left(\beta-\alpha\right)>0$ which implies $(M^2_Z-\lambda^2
v^2)\sin4\beta >  0$ for
$m_H>m_h$ when $\Delta_{\rm loop}$ is neglected. Note that, at tree level for $\tan\beta =1$, 
$\cot\left(\beta-\alpha\right)=0$ and the enhancement of the $ht\bar{t}$ coupling requires $\lambda v>(<) M_Z$ for $\tan\beta>(<)1$.  
This implies that $\tan\beta<1$ is disfavoured because the $tth$ enhancement is possible only if the tree-level Higgs mass is smaller than in MSSM
with
large $\tan\beta$, so (at least) one stop would have to be very heavy in order to account for 
the 125 GeV Higgs. Moreover, for $\tan\beta<1$ the top Yukawa coupling enters the non-perturbative regime close to the TeV scale.

\begin{figure}
 \includegraphics[width=0.49\textwidth]{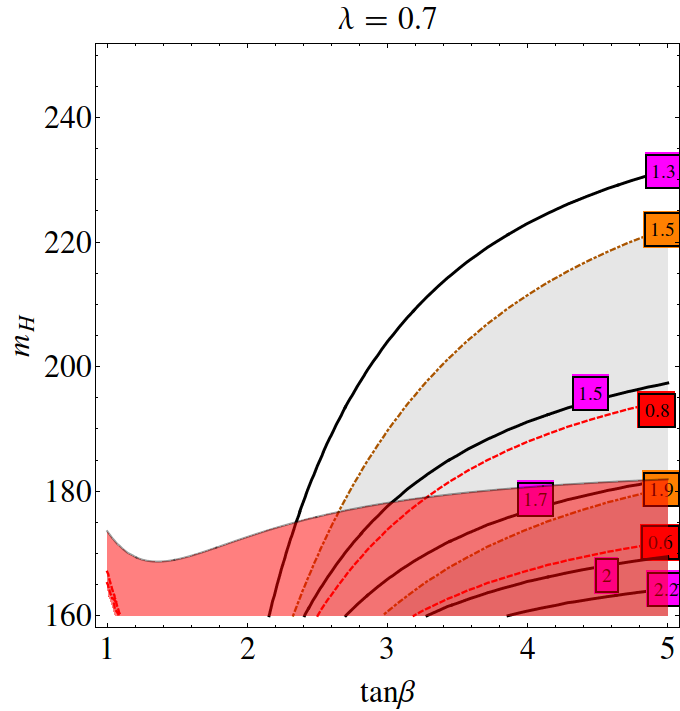}
  \includegraphics[width=0.49\textwidth]{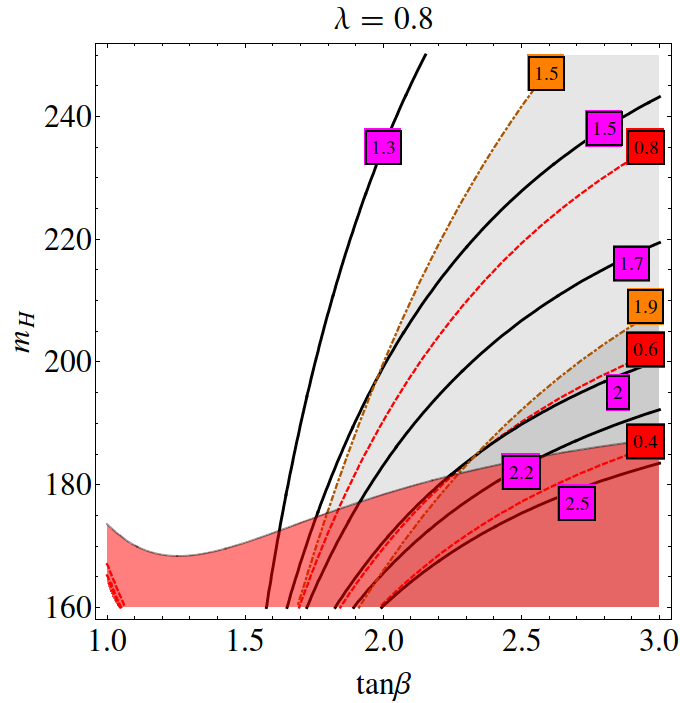}
     \includegraphics[width=0.49\textwidth]{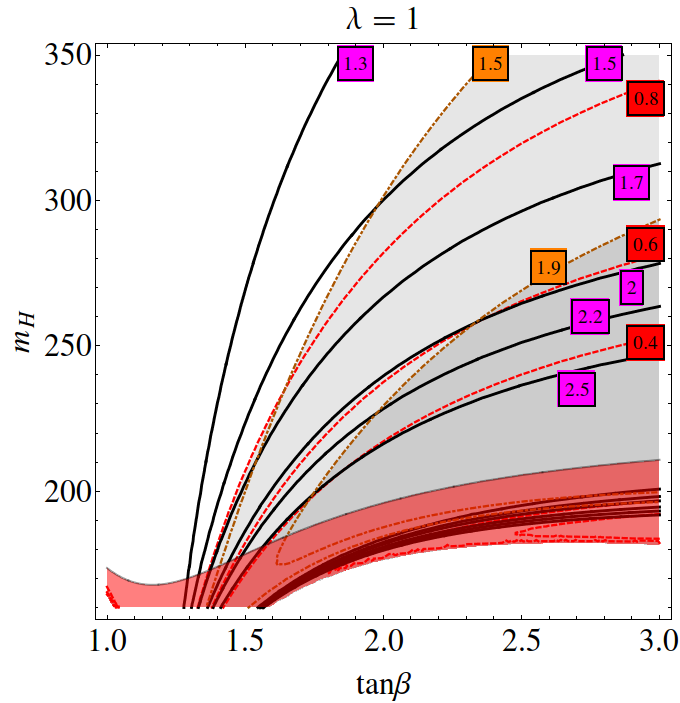}
   \includegraphics[width=0.49\textwidth]{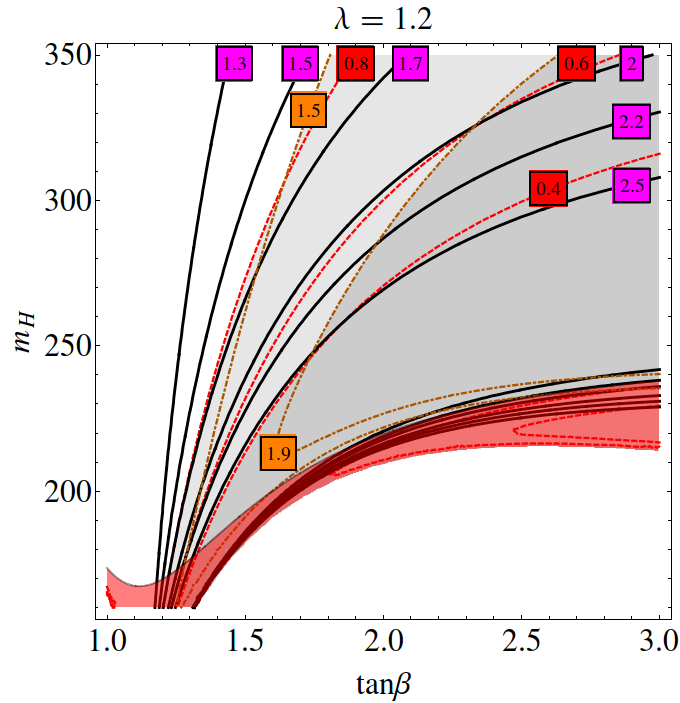}
\caption{The same as in Fig.~\ref{fig:tth_2HDMcontour} but for NMSSM in the ($\tan\beta$-$m_H$) plane for several values of $\lambda$ using the approximate
formula \eqref{eq:cot_betaalpha_NMSSM} with $\Delta_{\rm loop}=75$ GeV. The red shaded area is excluded because $m_{H^\pm}<160$ GeV there. The white area below
the red shaded area (visible in the lower panels) is theoretically inaccessible for $m_h=125$ GeV. }
\label{fig:NMSSMcontour}
\end{figure}

Notice also that $\Delta_{\rm loop}$, which is dominated by stop loops, is positive
\footnote{Negative $\Delta_{\rm loop}$ is possible only for very large stop mixing which would lead to destabilization of the EW vacuum
\cite{Vacuum_stop}.}
so after taking into account loop effects, for $\tan\beta>1$, the critical value of
$\lambda$, above which the $tth$ cross-section is enhanced, is larger than $M_Z/v$.
This may be easily understood by rewriting the expression of $\cot(\beta-\alpha)$ in terms
of  the $\hat{M}_{hh}^2$ matrix element, namely
\begin{equation}
\cot(\beta - \alpha) = -\frac{\hat{M}_{hh}^2 - \cos(2\beta) M_Z^2 - 2 \lambda^2 v^2 \sin^2(\beta)}{\tan\beta \left(\hat{M}_{HH}^2 - m_h^2\right)}
\end{equation}
Since $\hat{M}_{hh}^2 \simeq m_h^2$,  one can easily show that for $\tan\beta = {\cal{O}}(1)$ the lightest Higgs alignment, for which $\cot(\beta -\alpha) \simeq 0$, occur for values of $\lambda$ in the range $\lambda \simeq 0.65$--0.7~\cite{alignment}, with larger values of $\lambda$ leading to positive values of $\cot(\beta - \alpha)$ and hence to an enhancement of the top quark coupling to $h$. 

 In the rest of the presentation we fix $\Delta_{\rm
loop}=75$
GeV which 
is a typical value of the loop correction for the stop masses in the range of several hundreds GeV and large stop mixing. We checked that such value of
$\Delta_{\rm loop}$
for $\tan\beta\approx2$ leads to the results that are in a good agreement with a more precise calculation by {\tt NMSSMTools 4.8.1} (that diagonalizes
the full loop
corrected 3x3 NMSSM Higgs mass matrix) \cite{NMSSMTools} that shows that $\lambda\gtrsim0.6$ is required for the $tth$ enhancement. 
One technical comment is that we choose to fix $\Delta_{\rm loop}$ rather than adjust $\Delta_{\rm loop}$ to get the Higgs mass of 125 GeV. This is
because for large $\lambda$ that would require negative values of $\Delta_{\rm loop}$ which cannot be obtained if the vacuum stability constraints are
taken into account. We assume, instead,  that the Higgs mass is set to 125 GeV by mixing effects with the heavy singlet \cite{Hall_Higgs126}. Indeed, it can be
shown that the mixing with the singlet can give
large negative correction to $m_h$ even if this mixing changes the Higgs couplings in a negligible amount~\cite{nmssmmixing}.

In Fig.~\ref{fig:NMSSMcontour} contours of $R^{\rm tth}_{VV}$ in the plane ($\tan\beta$,$m_H$) for several values of $\lambda$ are  presented. In
these plots $m_h$ is fixed to 125 GeV and the Higgs couplings that enter the formulae for cross-sections and branching ratios are determined by
eq.~\eqref{eq:cot_betaalpha_NMSSM}. Notice that, in contrast to a general type-II 2HDM discussed in the previous section, values of $\tan\beta$ as
small as possible are no longer preffered. This is because in NMSSM
$\cot(\beta-\alpha)$ is not independent from $\tan\beta$ and as stressed above it actually vanishes at tree level in the limit $\tan\beta\to1$. In fact,
enhancement of  the $ht\bar{t}$ coupling (with respect to the Higgs coupling to massive gauge bosons) is maximized for $\tan\beta\approx2$. $R^{\rm
tth}_{VV}$
is maximized for even larger $\tan\beta$ due to larger suppression of the $hb\bar{b}$ coupling but as discussed above the latter possibility is constrained by the LHC data in
other channels. Therefore, after taking into account the experimental constraints $R^{\rm tth}_{VV}$ is typically maximal for $\tan\beta$ close to 2.

It can be also seen that if one demands  perturbativity up to the GUT scale, which for small $\tan\beta$ can be realised only for  $\lambda\lesssim0.7$,
substantial enhancement of 
$\sigma^{\rm tth}$ with respect to $\sigma^{\rm VBF/VH}$ is possible only for very light $H$. This is a consequence of the approximate alignment in
the NMSSM Higgs
sector for $\lambda\sim 0.6$~\cite{alignment}. However, the region of light $H$ is strongly constrained, because the CP-odd and charged Higgses are 
also light in such a case. At tree level:
\begin{align}
\label{mAeq}
m_A^2=\hat{M}^2_{HH}- (M^2_Z-\lambda^2 v^2)\sin^2(2\beta) \,, \\
\label{mHpmeq}
m_{H^{\pm}}^2=m_A^2+m_W^2-\lambda^2v^2 \,.
\end{align}
In the context of the MSSM, the constraints on light BSM Higgses were studied e.g. in Ref.~\cite{Djouadi_mAtanb}. However, the Higgs sector of NMSSM
with large $\lambda$ is significantly different from that of MSSM. Very important constraint
comes from the
charged Higgs searches. Particularly important search is in the channel $t\to H^{+}b$ ($H^+\to\tau^+\nu_{\tau}$) which for most values of $\tan\beta$  excludes
$m_{H^{\pm}}<160$ GeV both by ATLAS \cite{atlas_charged} and CMS \cite{cms_charged}. Slightly weaker bounds on $m_{H^{\pm}}$ have been found for $\tan\beta$ in
the range between 4 and 20 which has no big impact on our results since the $tth$ enhancement prefers lower values of $\tan\beta$. As can be seen from
Fig.~\ref{fig:NMSSMcontour}, this search excludes the smallest values of $m_H$ and the exclusion becomes stronger as $\lambda$ grows as a consequence
of relations \eqref{mAeq}-\eqref{mHpmeq}.
After taking this constraint into account the $ht\bar{t}$ coupling cannot be significantly enhanced for values of $\lambda$ consistent with the
perturbativity up to the GUT scale. In such a case $R^{\rm tth}_{VV}$ can only be enhanced as a result of the suppression  of the $hb\bar{b}$
coupling, that occurs at larger $\tan\beta$. For larger $\tan\beta$ perturbativity constraint on $\lambda$ becomes slightly weaker and can be satisifed e.g. for
$\lambda=0.76$ and $\tan\beta=4$. Such a case is represented by point P1 in Table \ref{tab:benchmarks} which consists a list of benchmarks obtained with
{\tt NMSSMTools}. One can see that $R^{\rm
VBF/VH}_{\gamma\gamma}$ for such $\tan\beta$ is always larger than $R^{\rm tth}_{VV}$ and provides the main constraint for the latter.

\begin{table}
\centering
\begin{tabular}{c|ccccc}
& {\rm P1} & {\rm P2} & {\rm P3} & {\rm P4} & {\rm P5}  \\
\hline
$\lambda$ & $0.76$ & $0.85$ & $1.1$   & $1.4$   & $1.4$ \\
$\tan \beta$ & 4 & 2 & 2 & 1.5 & 1.5\\
\hline
$m_{Q_3}$ & 700 & 700 & 700 & 700  & 700\\
$m_{U_3}$ & 500 & 480 & 500 & 480 & 450\\
$A_t$ & -1170 & -1100 & -1030 & -780 & -1030\\
\hline
$\mu$ & 300 & 770 & 1040 & 1060 & 390 \\
$M_2$ & 500 &  500 &  500 &  500 & -90 \\
$\mu'$ & 60 & 45 & 40 & 14 & -24\\
$M_{P_1}$ & 193 & 197 & 277 & 332 & 357\\
$M_{P_2}$ & 2000 & 2500 & 3000 & 2400 & 800\\
\hline
$m_{h}$ & 125.1 & 125.9 & 125.0 & 124.9 & 125.0 \\
$m_{H}$ & 192 & 184 & 262 &  280 & 299 \\
$m_{H^\pm}$ & 167 & 161 & 236 & 257 & 272  \\
$m_{A}$ & 195 & 204 & 293 & 342  & 344 \\
$m_{\tilde{\chi}_1^0}$ & 70 & 65 & 66 & 63 & 89  \\
$m_{\tilde{\chi}^\pm_1}$ & 282 & 504 & 516 & 514 & 109  \\
$m_{\tilde{t}_1}$ & 236 & 232 & 241 & 231 & 222  \\
$m_{\tilde{t}_2}$ & 726 & 752 & 766 & 757 & 730 \\
\hline
$R^{\rm tth}_{VV}$              & 1.79 & 1.84 & 1.96 & 1.92 & 1.87  \\
$R^{\rm tth}_{\gamma\gamma}$    & 1.97 & 2.12 & 2.22 & 2.19 & 1.96   \\
$R^{\rm gg}_{VV}$               & 1.16 & 1.00 & 1.12 & 1.18 & 1.23  \\
$R^{\rm gg}_{\gamma\gamma}$     & 1.29 & 1.15 & 1.27 & 1.34 & 1.29  \\
$R^{\rm VBF/VH}_{VV}$           & 1.70 & 1.57 & 1.65 & 1.48 & 1.43  \\
$R^{\rm VBF/VH}_{\gamma\gamma}$ & 1.89 & 1.80 & 1.87 & 1.69 & 1.50 \\
$R^{\rm VBF/VH}_{\tau\tau}$     & 0.70 & 0.71 & 0.67 & 0.71 & 0.65  \\
\hline
BR$(H\to\tilde{\chi}_1^0 \tilde{\chi}_1^0)$ & 0.71 & 0.49 & 0.24 & 0.14 & 0.19  \\
BR$(H\to \tilde{\chi}_1^0 \tilde{\chi}_2^0)$ & 0 & 0 & 0 & 0 & 0.17  \\
BR$(H\to hh)$ & 0 & 0 & 0.47 & 0.71 & 0.54  \\
BR$(A\to\tilde{\chi}_1^0 \tilde{\chi}_1^0)$ & 0.85 & 0.89 & 0.78 & 0.75 & 0.88  \\
BR$(A\to H^\pm W^\mp)$ & 0 & 0 & 0 & 0.05 & 0 \\
\end{tabular}
\caption{List of benchmark points obtained with {\tt NMSSMTools 4.8.1}. All masses are in GeV. All points satisfy all experimental constraints from
the Higgs signal strength
measurements, as well as from direct searches for Higgses, checked with {\tt HiggsBounds 4.2.1}
\cite{HiggsBounds}, and stops. The remaining soft sfermion masses are set to 2 TeV,
$M_3=1.5$ TeV,  $M_1=250$ GeV. All the remaining $A$-terms are set to $1.5$ TeV, while $\kappa=A_\kappa=0$. The remaining parameters are
calculated with {\tt NMSSMTools} using EWSB conditions and the values of $\mu$ and $M_{P_i}$ (with $M_{P_i}$ defined as the diagonal entries of the pseudoscalar mass matrix). The above spectra were obtained with the renormalization scale set to 700 GeV.}.
\label{tab:benchmarks}
\end{table}

Relaxing the requirement of perturbativity up to the GUT scale, substantial enhancement of the $ht\bar{t}$ coupling becomes possible resulting in $R^{\rm
tth}_{VV}\approx2$ without violating constraints from other Higgs signal strengths, as long as $\tan\beta$ is close to 2. Already for
$\lambda\approx0.8$ and $\tan\beta\approx2$, $R^{\rm tth}_{VV}\approx2$ can be obtained with $m_{H^{\pm}}>160$ GeV. However, there are
additional constraints coming from the LHC searches for the CP-even Higgs in the $ZZ$ and $WW$ decay channels \cite{ATLAS_WWhigh}-\cite{cms_WWZZ} and searches
for the CP-odd Higgs
in the $\tau\tau$ \cite{ATLAS_AHtautau,cms_AHtautau} and $hZ$ \cite{ATLAS_AtoZh,CMS_AtoZh} decay channels. Points with $R^{\rm tth}_{VV}\approx2$ typically
violate some of those constraints, especially the
constraint from the $H\to ZZ$ searches, unless $H$ and $A$ have significant fraction of invisible decays. Therefore, valid points with large
$tth$ enhancement must have light neutralino (but not lighter than $m_h/2$ to avoid invisible $h$ decays). Light neutralino is preferred also in order
to avoid the LHC constraints on light stop. Indeed, keeping the gluon fusion rates in the gauge boson decay channels close to the SM prediction when $R^{\rm
tth}_{VV}$ is enhanced requires the lightest stop mass to be below about 300 GeV \footnote{For larger $m_{\tilde{t}_1}$ the stop correction to the
effective Higgs coupling to gluons \eqref{eq:cg_stopeff} is too small unless $\tilde{X}_t/m_{\tilde{t}_2}$ is so large that the EW vacuum becomes unstable. }.
Such a light stop is excluded by the ATLAS \cite{ATLASstop1}-\cite{ATLASstop4}  and CMS \cite{CMSstop1}-\cite{CMSstop4} stop searches unless the mass
splitting between the stop and the LSP is
very close to the top mass, $W$ mass or zero. Moreover, for the stop mass below about 250 GeV the zero mass splitting between the stop and the LSP is excluded
by the CMS monojet search \cite{CMSstop3}. Therefore, generically if the light stop is consistent with the LHC data then some of the decays of the
heavy
Higgses are invisible. In the NMSSM with enhanced $tth$ rates, the best candidate for the LSP is singlino-like neutralino because due to the mixing with Higgsinos and the large values of $\lambda$, the
decay width of heavy Higgses to singlino is typically large (if kinematically accessible).

Points P2, P3 and P4 in Table \ref{tab:benchmarks} are the NMSSM points that have a Landau pole below the GUT scale and were obtained with {\tt NMSSMTools} and
satisfy all experimental constraints on
the Higgs sector, which was verified with {\tt HiggsBounds 4.2.1} \cite{HiggsBounds}. Constraints on the light stop are also satisfied because the
mass splitting between the stop and the LSP is very close to the top mass. All the benchmark points predict $R^{\rm tth}_{VV}\approx2$. Benchmark P2
is characterized by $\lambda=0.85$ and
$\tan\beta=2$ and $m_{H^{\pm}}$ just above 160 GeV. For smaller values of $\lambda$ and $\tan\beta=2$ we have not found points with $R^{\rm
tth}_{VV}\approx2$
that are
consistent simultanously with the LHC $H\to WW$ and $H\to ZZ$ searches. The crucial role for the benchmark P2 to be consistent with the Higgs
data is played by large branching ratios of $A$ and $H$ decays to pairs of LSP. 

Benchmark P3 with $\lambda=1.1$ and $\tan\beta=2$ is characterized by $m_H$ above $2 \ m_h$ and the main role in avoiding constraints from the $H\to
ZZ$ searches is played by large $\BR (H\to hh)$ but invisible decays are needed to avoid the constraints from $A\to h Z$ searches. For even
larger values of $\lambda$,  $R^{\rm tth}_{VV}\approx2$ can be obtained also for $\tan\beta$ significantly below 2. Such a case is represented by
benchmark P4 with $\lambda=1.4$ and $\tan\beta=1.5$. Notice that in this case $R^{\rm tth}_{VV}$ is similar to other benchmarks but the VBF rates are
smaller than for $\tan\beta=2$. Note also that for such a large $\lambda$ the splitting between the charged Higgs mass and
CP-odd Higgs mass is so large that decays of the latter to the charged Higgs and W boson become kinematically accessible which additionally helps in
satisfying the constraints from $A\to hZ$ searches.

It should be noted  that  $R^{\rm tth}_{VV}$ of about 2 in the NMSSM typically ruins the $1\sigma$ agreement with the combined VBF measurements in the
$\gamma\gamma$ decay channel because although CMS observed an enhancement, ATLAS observed some suppression (with respect to the SM prediction) in this channel. This feature is specific to
NMSSM and results from the approach to alignment in the limit $\tan\beta\to1$. As emphasized before, in general type-II 2HDM (with new colored states that keep the gluon fusion
production rate close to the SM prediction) $R^{\rm tth}_{VV}$ of about 2 is possible without large modifications to the VBF rates provided that
$\tan\beta\approx1$, cf. benchmark B1 in Table~\ref{tab:benchmarks2HDM}. 

Nevertheless,  strongly enhanced $R^{\rm tth}_{VV}$ without violating $1\sigma$ agreement with the combined VBF measurements in the $\gamma\gamma$
decay channel can also be obtained in the NMSSM provided that a chargino is very light and ${\rm sgn} (\mu M_2)<0$. In such a case the chargino
loop contribution to the $\gamma\gamma$ decay rate interferes destructively with the dominant W boson loop. In order to substantially alter the
$\gamma\gamma$ rate the lightest chargino should be not far above 100 GeV, which is a generic lower mass limit for chargino from LEP
\cite{charginoLEP}, with non-negligible mixing between higgsino and gaugino component \cite{Casas:2013pta,Batell:2013bka}.\footnote{For large
$\lambda$ the $\gamma\gamma$ rate can be also modified if higgsino-dominated chargino is light and the Higgs has a non-negligible singlet component
\cite{diphoton_chargino,ChoiNMSSM}.} This effect is demonstrated by benchmark P5 in Table \ref{tab:benchmarks} where $R^{\rm tth}_{VV}$ of about 1.9
is obtained with $R^{\rm VBF/VH}_{\gamma\gamma}\approx 1.5$.
For benchmark P5, the stop collider phenomenology differs from other benchmarks because the lightest stop can decay to the lightest chargino and a
bottom quark. In such a case limits for direct stop production typically become stronger, but some parts of parameter space with light stop are still
allowed. For example, a stop with mass of 220 GeV decaying to a chargino and a bottom quark in the case of  a~20 GeV mass splitting between the chargino and
the LSP, with the LSP mass around 90 GeV, as it is the case for benchmark P5, is consistent with the LHC data \cite{ATLASstop1,ATLASstop4,CMSstop1}. Due to the presence of a 
light wino-dominated chargino in benchmark 5, limits for direct wino-like $\tilde{\chi}^\pm_1-\tilde{\chi}_2^0$ production may also be relevant.  In this case, 
$\tilde{\chi}^\pm_1$ decays to a W boson and the LSP and the mass limits for $\tilde{\chi}^\pm_1$ (assumed to be degenerate with $\tilde{\chi}_2^0$)
depend on the decay pattern of $\tilde{\chi}_2^0$. For benchmark 5, $\tilde{\chi}_2^0$ decays to the LSP and a photon or off-shell $Z$
boson with BR$(\tilde{\chi}_2^0 \to \tilde{\chi}_1^0 \gamma)\approx 55\%$. For both decay patterns the LHC searches are not yet sensitive for such a
small mass splitting between the chargino and the LSP~\cite{chargino_photon,charginoZ_ATLAS,charginoZ_CMS}. 

Let us also comment on the fact that benchmark points P2-P5 in Table \ref{tab:benchmarks} are in conflict with B-physics constraints if minimal
flavour violation (MFV) is assumed. 
In particular ${\rm BR}(b \to s \gamma)$ is typically about $5\cdot10^{-4}$ which is somewhat above the experimental value \cite{bsg_exp}. This
tension originates from large loop contributions from light highly-mixed stops and the charged Higgs. Nevertheless, ${\rm BR}(b \to s \gamma)$ can be
brought in agreement with the experimental data by arranging parameters such that the charged Higgs contribution to ${\rm BR}(b \to s \gamma)$ is
approximately canceled by the corresponding stop contribution. One should also keep in mind that B-physics observables are sensitive to flavour
structure of the down squark parameters via loops with gluinos so they can be brought in agreement with measurements by adjusting non-MFV parameters
\cite{Gabbiani:1996hi}.

\section{Conclusions}
\label{sec:concl}

We have investigated enhancement of the $tth$ production cross-section in models with the Higgs sector that can be approximately described as type-II
2HDM. We have shown that in this class of models the $tth$ signal rates in the gauge boson decay channels can be more than two times larger than in the
SM, as hinted by the ATLAS and CMS excesses, provided that $\tan\beta$ is small and additional light colored particles, such as the stop,
interfere destructively with the top quark in the gluon fusion amplitude.   In these models, the necessary decrease of the top quark coupling to the lightest Higgs is associated with a reduction of the bottom quark coupling, which contributes to an enhancement of the Higgs decay into gauge bosons. 

We have also shown that large $tth$ enhancement of about two can be realized in the NMSSM, 
although the situation is more constrained in this case, due to the specific dependence of the CP-even Higgs matrix elements on the model parameters. 
For instance, this requires  values of $\lambda$ larger than the ones allowing the perturbative consistency of the theory up to the GUT scale.
Moreover, $\tan\beta$ must be above one (preferably between 1.5 and 2), what implies a sizable reduction of the bottom coupling to the lightest Higgs boson and
hence a large enhancement of the decays of the lightest CP-even Higgs into gauge bosons. It should be noted that the NMSSM realization of $tth$ enhancement is not generic and requires some tuning in the stop sector to keep the gluon fusion rates close to the SM prediction. Moreover, since this scenario points to large values of $\lambda$ and small $\tan\beta$ the Higgs mass generically turns out to be too large but can be set to 125 GeV by introducing small amount of mixing between the Higgs and the singlet scalar which partially cancels large contribution to the Higgs mass proportional to $\lambda$.

If the $tth$ excess persists in the LHC run 2 data, the NMSSM interpretation of it can be tested at the LHC in multiple ways. First of all, since
signal rates in VBF production channel show correlated deviations with the $tth$ signal rates, improved measurements of the VBF production mode,
especially in the $\gamma\gamma$ decay channel, can set strong constraints on this scenario. Secondly, the gluon fusion signal strengths are close to
the SM prediction due to the presence of a light stop with mass below 300 GeV, which is consistent with current LHC searches because its mass splitting with the LSP is
close to the top quark mass, or because there is an additional light chargino with mass close to 100~GeV and a few tens of GeV heavier than the LSP.  Therefore, direct stop (and in some scenarios chargino) searches in this region of parameters can also efficiently probe this model. Light stop contribution to the gluon fusion cross-section can be also probed by looking for a boosted Higgs with a jet \cite{Grojean:2013nya}. Finally, this scenario can be tested at the LHC by direct searches of MSSM-like Higgs bosons which masses have to be in the range of several hundred of GeV to allow for substantial $tth$ enhancement.

\section*{Acknowledgments}
This work has been partially supported by National Science Centre under research grant DEC-2014/15/B/ST2/02157. MB acknowledges support from the
Polish 
Ministry of Science and Higher Education (decision no.\ 1266/MOB/IV/2015/0).
 Work at ANL is supported in part by the U.S. Department of Energy, Office of High Energy Physics, under Contract No. DE-AC02-06CH11357.
Work at the University of Chicago is supported in part by U.S. Department of Energy grant number DE-FG02-13ER41958. MB would like to thank  Cyril Hugonie, Ulrich Ellwanger and Tim Stefaniak for useful correspondence about {\tt NMSSMTools} and {\tt HiggsBounds}.
MB thanks the Galileo Galilei Institute for Theoretical Physics and INFN for hospitality and partial support during the completion of this work. C.W thanks the 
hospitality of the Aspen Center for Physics, which is supported by the National Science Foundation under  Grant No. PHYS-1066293. 

\section*{Appendix}

In the computation of cross-sections and branching ratios normalized to the SM values  we use the formalism of Ref.~\cite{Falkowski}.
In 2HDM, deviations from the SM predictions occur through the modifications of the Yukawa
coupling to up-type fermions, $c_t$, the Yukawa coupling to down-type fermions, $c_b$, and the couplings to $W$ and $Z$ bosons, $c_V$, which are normalised to
the SM values. Using these normalised couplings the most relevant Higgs decay widths  are given by:
\begin{align}
 &\Gamma(h\to VV) = c_V^2 \Gamma(h\to VV )^{\rm SM} \,, \\
  &\Gamma(h\to bb/\tau\tau) = c_b^2 \Gamma(h\to bb/\tau\tau )^{\rm SM} \,, \\
   &\Gamma(h\to cc) = c_t^2 \Gamma(h\to cc )^{\rm SM} \,, \\
  &\Gamma(h\to gg) = \left|\frac{\hat{c}_g}{\hat{c}_g^{\rm SM}}\right|^2 \Gamma(h\to gg )^{\rm SM} \,, \\
    &\Gamma(h\to \gamma\gamma) = \left|\frac{\hat{c}_{\gamma}}{\hat{c}_{\gamma}^{\rm SM}}\right|^2 \Gamma(h\to gg )^{\rm SM} \,.
\end{align}  
The decays to gluons and photons are loop-induced and the leading contribution to these decays can be described by dimension-5 operators with  $\hat{c}_g$ and
$\hat{c}_{\gamma}$ being the corresponding effective Higgs couplings to gluons and photons, respectively, which are approximately given by:
\begin{equation}
\label{eq:hatcg}
 \hat{c}_g=c_g+(-0.06+0.09i)c_b \,, \qquad \hat{c}_{\gamma}=c_{\gamma} - 1.04c_V \,.
\end{equation}
The SM values of $c_g$ and $c_{\gamma}$, which arise from integrating out a top quark, are approximately given by:
\begin{align}
 c_g^{SM}\approx1.03   \,, \\
 c_{\gamma}^{SM}\approx\frac{2}{9}\, 1.03   \,. 
\end{align}
Beyond the SM, $c_g$ and $c_{\gamma}$ are given by:
\begin{align}
\label{eq:cgBSM}
 c_g=c_g^{SM} c_t + \delta c_g   \,, \\
 \label{eq:cgamBSM}
 c_{\gamma}=c_{\gamma}^{SM} c_t + \delta c_{\gamma}   \,. 
\end{align}
where $\delta c_i$ stand for the contributions from new particles that couple to the Higgs.
 
The production cross-sections scale like:
\begin{align}
&\sigma^{tth}\equiv \frac{\sigma(gg\to tth)}{\sigma^{SM}(gg\to tth)}=c_t^2 \,, \\
&\sigma^{gg}\equiv \frac{\sigma(gg\to h)}{\sigma^{SM}(gg\to h)}=\left|\frac{\hat{c}_g}{\hat{c}_g^{\rm SM}} \right|^2 \,, \\
&\sigma^{VBF}\equiv \frac{\sigma(q\bar{q}\to hjj)}{\sigma^{SM}(q\bar{q}\to hjj)} = \sigma^{VH}\equiv  \frac{\sigma(q\bar{q}\to hV)}{\sigma^{SM}(q\bar{q}\to
hV)}=c_V^2 \,, \\
\end{align}



\begin{thebibliography}{99}

\bibitem{Higgsdiscovery}
  G.~Aad {\it et al.}  [ATLAS Collaboration],
  Phys.\ Lett.\ B {\bf 716} (2012) 1
  [arXiv:1207.7214 [hep-ex]];
  S.~Chatrchyan {\it et al.}  [CMS Collaboration],
  Phys.\ Lett.\ B {\bf 716} (2012) 30
  [arXiv:1207.7235 [hep-ex]].

\bibitem{Higgsdata}
  The ATLAS and CMS Collaborations,
  ATLAS-CONF-2015-044,  CMS-PAS-HIG-15-002. 

\bibitem{gluonfusionSM}
C.~Anastasiou, C.~Duhr, F.~Dulat, E.~Furlan, T.~Gehrmann, F.~Herzog, A.~Lazopoulos and B.~Mistlberger,
  arXiv:1602.00695 [hep-ph];
D.~de Florian and M.~Grazzini,
  Phys.\ Lett.\ B {\bf 674}, 291 (2009)
  doi:10.1016/j.physletb.2009.03.033
  [arXiv:0901.2427 [hep-ph]];
S.~Alioli, P.~Nason, C.~Oleari and E.~Re,
  JHEP {\bf 0904}, 002 (2009)
  doi:10.1088/1126-6708/2009/04/002
  [arXiv:0812.0578 [hep-ph]];
C.~Anastasiou, R.~Boughezal and F.~Petriello,
  JHEP {\bf 0904}, 003 (2009)
  doi:10.1088/1126-6708/2009/04/003
  [arXiv:0811.3458 [hep-ph]];
V.~Ahrens, T.~Becher, M.~Neubert and L.~L.~Yang,
  Eur.\ Phys.\ J.\ C {\bf 62}, 333 (2009)
  doi:10.1140/epjc/s10052-009-1030-2
  [arXiv:0809.4283 [hep-ph]].
  
\bibitem{Azatov:2013xha}
  A.~Azatov and A.~Paul,
  JHEP {\bf 1401} (2014) 014
  doi:10.1007/JHEP01(2014)014
  [arXiv:1309.5273 [hep-ph]].
  
\bibitem{Grojean:2013nya}
  C.~Grojean, E.~Salvioni, M.~Schlaffer and A.~Weiler,
  JHEP {\bf 1405} (2014) 022
  doi:10.1007/JHEP05(2014)022
  [arXiv:1312.3317 [hep-ph]].
  
\bibitem{tthATLAS}
  G.~Aad {\it et al.} [ATLAS Collaboration],
  Phys.\ Lett.\ B {\bf 749} (2015) 519
  [arXiv:1506.05988 [hep-ex]];
  G.~Aad {\it et al.} [ATLAS Collaboration],
  Phys.\ Lett.\ B {\bf 740} (2015) 222
  [arXiv:1409.3122 [hep-ex]].
  
\bibitem{tthCMS}
  V.~Khachatryan {\it et al.} [CMS Collaboration],
  JHEP {\bf 1409} (2014) 087
   [JHEP {\bf 1410} (2014) 106]
  [arXiv:1408.1682 [hep-ex]].

\bibitem{Huang:2015fba}
  P.~Huang, A.~Ismail, I.~Low and C.~E.~M.~Wagner,
  Phys.\ Rev.\ D {\bf 92} (2015) 7,  075035
  doi:10.1103/PhysRevD.92.075035
  [arXiv:1507.01601 [hep-ph]].
  
\bibitem{Chen:2015jmn}
  C.~R.~Chen, H.~C.~Cheng and I.~Low,
  arXiv:1511.01452 [hep-ph].  
  

  
\bibitem{Angelescu:2015kga}
  A.~Angelescu, A.~Djouadi and G.~Moreau,
  arXiv:1510.07527 [hep-ph].

\bibitem{reviewEllwanger}
  U.~Ellwanger, C.~Hugonie and A.~M.~Teixeira,
  Phys.\ Rept.\  {\bf 496} (2010) 1
  [arXiv:0910.1785 [hep-ph]].
  
\bibitem{Falkowski}
  D.~Carmi, A.~Falkowski, E.~Kuflik, T.~Volansky and J.~Zupan,
  JHEP {\bf 1210} (2012) 196
  doi:10.1007/JHEP10(2012)196
  [arXiv:1207.1718 [hep-ph]].

\bibitem{Djouadi}
  A.~Djouadi,
  Phys.\ Rept.\  {\bf 459} (2008) 1
  [hep-ph/0503173].
  
\bibitem{NSUSYfits}
  J.~R.~Espinosa, C.~Grojean, V.~Sanz and M.~Trott,
  JHEP {\bf 1212} (2012) 077
  [arXiv:1207.7355 [hep-ph]].


  
  \bibitem{nmssmmixing}
  M.~Badziak, M.~Olechowski and S.~Pokorski,
  JHEP {\bf 1306} (2013) 043
  [arXiv:1304.5437 [hep-ph]].
 
 \bibitem{alignment}
  M.~Carena, I.~Low, N.~R.~Shah and C.~E.~M.~Wagner,
  JHEP {\bf 1404} (2014) 015
  [arXiv:1310.2248 [hep-ph]];
  M.~Carena, H.~E.~Haber, I.~Low, N.~R.~Shah and C.~E.~M.~Wagner,
  arXiv:1510.09137 [hep-ph].

\bibitem{Vacuum_stop}
  D.~Chowdhury, R.~M.~Godbole, K.~A.~Mohan and S.~K.~Vempati,
  JHEP {\bf 1402} (2014) 110
  [arXiv:1310.1932 [hep-ph]];
    N.~Blinov and D.~E.~Morrissey,
  JHEP {\bf 1403} (2014) 106
  [arXiv:1310.4174 [hep-ph]];
  J.~E.~Camargo-Molina, B.~Garbrecht, B.~O'Leary, W.~Porod and F.~Staub,
  Phys.\ Lett.\ B {\bf 737} (2014) 156
  [arXiv:1405.7376 [hep-ph]].
  
\bibitem{NMSSMTools}
  U.~Ellwanger, J.~F.~Gunion and C.~Hugonie,
  JHEP {\bf 0502} (2005) 066
  doi:10.1088/1126-6708/2005/02/066
  [hep-ph/0406215];
  U.~Ellwanger and C.~Hugonie,
  Comput.\ Phys.\ Commun.\  {\bf 175} (2006) 290
  doi:10.1016/j.cpc.2006.04.004
  [hep-ph/0508022].


  
\bibitem{Hall_Higgs126}
  L.~J.~Hall, D.~Pinner and J.~T.~Ruderman,
  JHEP {\bf 1204} (2012) 131
  [arXiv:1112.2703 [hep-ph]].
  
  \bibitem{Djouadi_mAtanb}
  A.~Djouadi, L.~Maiani, A.~Polosa, J.~Quevillon and V.~Riquer,
  JHEP {\bf 1506} (2015) 168
  [arXiv:1502.05653 [hep-ph]].
 



    
  \bibitem{atlas_charged}
  The ATLAS collaboration [ATLAS Collaboration],
  ATLAS-CONF-2013-090.
  
  \bibitem{cms_charged}
  V.~Khachatryan {\it et al.} [CMS Collaboration],
  JHEP {\bf 1511} (2015) 018
  doi:10.1007/JHEP11(2015)018
  [arXiv:1508.07774 [hep-ex]].

\bibitem{HiggsBounds}
  P.~Bechtle, O.~Brein, S.~Heinemeyer, O.~Stål, T.~Stefaniak, G.~Weiglein and K.~E.~Williams,
  Eur.\ Phys.\ J.\ C {\bf 74} (2014) 3,  2693
  doi:10.1140/epjc/s10052-013-2693-2
  [arXiv:1311.0055 [hep-ph]].
  
\bibitem{ATLAS_WWhigh}
  G.~Aad {\it et al.} [ATLAS Collaboration],
  JHEP {\bf 1601} (2016) 032
  doi:10.1007/JHEP01(2016)032
  [arXiv:1509.00389 [hep-ex]].

\bibitem{ATLAS_WWlow}
  G.~Aad {\it et al.} [ATLAS Collaboration],
  Phys.\ Rev.\ D {\bf 92} (2015) 1,  012006
  doi:10.1103/PhysRevD.92.012006
  [arXiv:1412.2641 [hep-ex]].
  
\bibitem{ATLAS_ZZ}
  G.~Aad {\it et al.} [ATLAS Collaboration],
detector,''
  Eur.\ Phys.\ J.\ C {\bf 76} (2016) 1,  45
  doi:10.1140/epjc/s10052-015-3820-z
  [arXiv:1507.05930 [hep-ex]].
  
\bibitem{cms_WW}
  S.~Chatrchyan {\it et al.} [CMS Collaboration],
  JHEP {\bf 1401} (2014) 096
  doi:10.1007/JHEP01(2014)096
  [arXiv:1312.1129 [hep-ex]].
  
  \bibitem{cms_ZZ}
  S.~Chatrchyan {\it et al.} [CMS Collaboration],
  Phys.\ Rev.\ D {\bf 89} (2014) 9,  092007
  [arXiv:1312.5353 [hep-ex]].
  
\bibitem{cms_WWZZ}
  V.~Khachatryan {\it et al.} [CMS Collaboration],
  JHEP {\bf 1510} (2015) 144
  doi:10.1007/JHEP10(2015)144
  [arXiv:1504.00936 [hep-ex]].
  
  
  \bibitem{ATLAS_AHtautau}
  G.~Aad {\it et al.} [ATLAS Collaboration],
  JHEP {\bf 1411} (2014) 056
  [arXiv:1409.6064 [hep-ex]].
  
  \bibitem{cms_AHtautau}
  V.~Khachatryan {\it et al.} [CMS Collaboration],
  JHEP {\bf 1410} (2014) 160
  [arXiv:1408.3316 [hep-ex]].
  
  \bibitem{ATLAS_AtoZh}
  G.~Aad {\it et al.} [ATLAS Collaboration],
  Phys.\ Lett.\ B {\bf 744} (2015) 163
  [arXiv:1502.04478 [hep-ex]].
  
  \bibitem{CMS_AtoZh}
 V.~Khachatryan {\it et al.} [CMS Collaboration],
  Phys.\ Lett.\ B {\bf 748} (2015) 221
  doi:10.1016/j.physletb.2015.07.010
  [arXiv:1504.04710 [hep-ex]].







\bibitem{ATLASstop1}
  G.~Aad {\it et al.} [ATLAS Collaboration],
  JHEP {\bf 1411} (2014) 118
  doi:10.1007/JHEP11(2014)118
  [arXiv:1407.0583 [hep-ex]].
  
\bibitem{ATLASstop2}
  G.~Aad {\it et al.} [ATLAS Collaboration],
  Phys.\ Rev.\ D {\bf 90} (2014) 5,  052008
  doi:10.1103/PhysRevD.90.052008
  [arXiv:1407.0608 [hep-ex]].
  
\bibitem{ATLASstop3}
  G.~Aad {\it et al.} [ATLAS Collaboration],
  Eur.\ Phys.\ J.\ C {\bf 75} (2015) 10,  510
  doi:10.1140/epjc/s10052-015-3726-9
  [arXiv:1506.08616 [hep-ex]].
  
\bibitem{ATLASstop4}
  G.~Aad {\it et al.} [ATLAS Collaboration],
  JHEP {\bf 1310} (2013) 189
  doi:10.1007/JHEP10(2013)189
  [arXiv:1308.2631 [hep-ex]].
  
\bibitem{CMSstop1}
  S.~Chatrchyan {\it et al.} [CMS Collaboration],
  Eur.\ Phys.\ J.\ C {\bf 73} (2013) 12,  2677
  doi:10.1140/epjc/s10052-013-2677-2
  [arXiv:1308.1586 [hep-ex]].
  
\bibitem{CMSstop2}
  S.~Chatrchyan {\it et al.} [CMS Collaboration],
  Phys.\ Lett.\ B {\bf 733} (2014) 328
  doi:10.1016/j.physletb.2014.04.023
  [arXiv:1311.4937 [hep-ex]].
  
\bibitem{CMSstop3}
  V.~Khachatryan {\it et al.} [CMS Collaboration],
  JHEP {\bf 1506} (2015) 116
  doi:10.1007/JHEP06(2015)116
  [arXiv:1503.08037 [hep-ex]].
  
\bibitem{CMSstop4}
  V.~Khachatryan {\it et al.} [CMS Collaboration],
sqrt(s) = 8 TeV,''
  arXiv:1512.08002 [hep-ex].


  
\bibitem{charginoLEP}
  G.~Abbiendi {\it et al.} [OPAL Collaboration],
  Phys.\ Lett.\ B {\bf 572} (2003) 8
  doi:10.1016/S0370-2693(03)00639-7
  [hep-ex/0305031].
  
\bibitem{Casas:2013pta}
  J.~A.~Casas, J.~M.~Moreno, K.~Rolbiecki and B.~Zaldivar,
  JHEP {\bf 1309} (2013) 099
  doi:10.1007/JHEP09(2013)099
  [arXiv:1305.3274 [hep-ph]].
  
\bibitem{Batell:2013bka}
  B.~Batell, S.~Jung and C.~E.~M.~Wagner,
  JHEP {\bf 1312} (2013) 075
  doi:10.1007/JHEP12(2013)075
  [arXiv:1309.2297 [hep-ph]].
  
\bibitem{diphoton_chargino}
  K.~Schmidt-Hoberg and F.~Staub,
  JHEP {\bf 1210} (2012) 195
  [arXiv:1208.1683 [hep-ph]].
  
\bibitem{ChoiNMSSM}
  K.~Choi, S.~H.~Im, K.~S.~Jeong and M.~Yamaguchi,
  JHEP {\bf 1302} (2013) 090
  [arXiv:1211.0875 [hep-ph]].
  
\bibitem{chargino_photon}
  G.~Aad {\it et al.} [ATLAS Collaboration],
  Phys.\ Rev.\ D {\bf 92} (2015) 7,  072001
  doi:10.1103/PhysRevD.92.072001
  [arXiv:1507.05493 [hep-ex]].

\bibitem{charginoZ_ATLAS}
  G.~Aad {\it et al.} [ATLAS Collaboration],
  JHEP {\bf 1405} (2014) 071
  doi:10.1007/JHEP05(2014)071
  [arXiv:1403.5294 [hep-ex]].
  
\bibitem{charginoZ_CMS}
  V.~Khachatryan {\it et al.} [CMS Collaboration],
  Eur.\ Phys.\ J.\ C {\bf 74} (2014) 9,  3036
  doi:10.1140/epjc/s10052-014-3036-7
  [arXiv:1405.7570 [hep-ex]].
  
  \bibitem{bsg_exp}
  Y.~Amhis {\it et al.} [Heavy Flavor Averaging Group (HFAG) Collaboration],
  arXiv:1412.7515 [hep-ex].
  
  \bibitem{Gabbiani:1996hi}
  F.~Gabbiani, E.~Gabrielli, A.~Masiero and L.~Silvestrini,
  Nucl.\ Phys.\ B {\bf 477} (1996) 321
  doi:10.1016/0550-3213(96)00390-2
  [hep-ph/9604387].
  


\end{thebibliography}
\end{document}